%Paper: hep-th/9408156
%From: Hideaki AOYAMA <aoyama@phys.h.kyoto-u.ac.jp>
%Date: Fri, 26 Aug 1994 17:37:16 +0900

%%%%%%%%%%%%%%%%%%%%%%%%%%%%%%%%%%%%%%%%%%%%%%%%%%%%%%%%%%%%%%%%%%%%%%%%%%%%%%%
%                                                                             %
%                         Bounce in Valley:                                   %
%                  Study of the extended structures                           %
%             from thick-wall to thin-wall vacuum bubbles                     %
%                                                                             %
%                    Hideaki Aoyama and Shinya Wada                           %
%                                                                             %
%                  (Plain TeX + phyzzx macro: 9 pages)                        %
%              (tar-gzipped-uuencoded postscript: 4 figures)                  %
%                                                                             %
%%%%%%%%%%%%%%%%%%%%%%%%%%%%%%%%%%%%%%%%%%%%%%%%%%%%%%%%%%%%%%%%%%%%%%%%%%%%%%%

\input phyzzx.tex   
\def\kyoto{\centerline{\sl Department of Fundamental Sciences}
	\centerline{\sl Faculty of Integrated Human Studies} 
          \centerline{\sl Kyoto University}
          \centerline{\sl Yoshida, Kyoto 606-01, Japan}} 
\def\NP{Nucl.~Phys.~}
\def\PR{Phys.~Rev.~}
\def\PRL{Phys.~Rev.~Lett.~}
\def\PL{Phys.~Lett.~}

% My title page modifications 
\catcode`\@=11 % This allows us to modify PLAIN macros.
\def\author#1{\vskip\frontpageskip\titlestyle{\fourteencp #1}\nobreak}
\def\abstract{\par\dimen@=\prevdepth \hrule height\z@ \prevdepth=\dimen@
   \vskip\frontpageskip\centerline{\twelverm ABSTRACT}\vskip\headskip }
\def\title#1{\vskip\frontpageskip \titlestyle{\seventeenbf #1} \vskip\headskip }\catcode`\@=12 % at signs are no longer letters

\VOFFSET = 1.2cm
\HOFFSET = .7cm
\pubnum{KUCP-0069}
\date{August 1994}
\titlepage
\title{\doublespace Bounce in Valley:\break
Study of the extended structures from thick-wall to thin-wall vacuum bubbles}
\author{Hideaki Aoyama
\foot{E-mail address: aoyama@phys.h.kyoto-u.ac.jp}
and 
Shinya Wada
\foot{E-mail address: shinya@phys.h.kyoto-u.ac.jp}}

\vskip 0.3cm
\kyoto
\midinsert\narrower\narrower
\abstract
\noindent
The valley structure associated with quantum meta-stability is examined.
It is defined by the new valley equation, which enables consistent 
evaluation of the imaginary-time path-integral.
We study the structure of this new valley equation and
solve these equations numerically.
The valley is shown to contain the bounce solution, as well as
other bubble structures.
We find that even when the bubble solution has thick wall, 
the outer region of the valley is made of large-radius, thin-wall bubble,
which interior is occupied by the true-vacuum.
Smaller size bubbles, which contribute to decay 
at higher energies, are also identified.
\endinsert

\endpage
%%%%%%%%%%%%%%%%%%%%%%%%%%%%%%%%%%%%%%%%%%%%%%%%%%%%%%%%%%%%%%%%%%%%%%%%
\normalspace

The decay of the false vacuum can be treated 
in the imaginary-time path-integral formalism
using the bounce solution.\REFS
\coleman{S.~Coleman \journal\PR &D15 (77) 2929.}
\REFSCON\callancoleman{C.~Callan and S.~Coleman \journal \PR &D16 (77) 1762.}
\refsend
Due to the existence of a negative-eigenvalue fluctuation-mode
around the bounce solution, 
the contour of the gaussian integration through the bounce
has to be deformed to yield the imaginary part of the energy level.
Thus the decay rate of the false vacuum is obtained.

This bounce belongs to a valley of the
action in the whole functional space.
This situation is analogous to that of 
the tunneling processes via instanton, 
such as the baryon number violation process in the standard model.
\REFS
\ring{A. Ringwald \journal \NP &B330 (90) 1.}
\REFSCON\esp{O. Espinosa  \journal \NP &B343 (90) 310.}
\REFSCON\allak{
H.~Aoyama and H.~Kikuchi \journal \PL &247B (90) 75;
\PR {\bf 43} (1991), 1999;
{\sl Int.~J.~of Mod.~Phys.} {\bf A7} (1992), 2741.}
\refsend

In either cases, when the initial condition is such that the 
state is in local minimum, the solutions of equations of motion
(bounce or instanton) dominates the relevant imaginary-time
path integral.
However, when the initial state is of higher energy,
different configurations could dominate.
For the quantum tunneling, the deformed instanton and anti-instanton
pair is known to play that role.\Ref\mattis{P.~B.~Arnold and M.~P.Mattis
\journal \PRL &66 (91) 13.}
These configuration generally belong to a valley of action,
since they form a line of relatively small actions.
Initially, streamline method was used to define the valley.\REFS
\shuryak{E.V.Shuryak \journal \NP &B302 (88) 559; {\sl ibid.} 621.}
\REFSCON\balitsky{I.~I.~Balitzky and A.~V.~Yung \journal \PL &168B (86) 113;
\NP \nextline {\bf B274} (1986), 475.}
\REFSCON\khoze{V.~V.~Khoze and A.~Ringwald \journal \PL &B259 (91) 106.}
\refsend
Later, one of the authors (H.A.) and Kikuchi proposed
an alternative definition of the valley, 
the {\sl new valley method}.\Ref
\newvalley{H.~Aoyama and H.~Kikuchi \journal\NP &B369 (92) 219.}
This is obtained by
separation of a collective coordinate that corresponds small, zero or 
negative eigenvalue, which is dangerous for the gaussian integration.
In this letter, we employ this new valley method to 
study the structure of the valley that contains the bounce.

We consider a quantum field theory of a scalar field $\phi(x)$ in 
3+1 dimensional space-time with action S, which we shall specify later.
The new valley equation is given by the following,
$$\int d^4x^\prime D(x, x^\prime)
		{\delta S \over \delta \phi(x^\prime)} = 
	\lambda {\delta S \over \delta\phi(x)}, \quad 
D(x, x^\prime) \equiv {\delta^2 S \over \delta \phi(x) \delta \phi(x^\prime)}.
\eqn\newvalley$$ 
In the above, the parameter $\lambda$ is the smallest eigenvalue of the
second-order differential operator $D(x, x^\prime)$.
In general, the new valley equation \newvalley\ is 
a fourth-order differential equation for $\phi(x)$.
We find it convenient to introduce an auxiliary field
to rewrite \newvalley\ to a set of
two second-order differential equations.
This is done in the following manner:
The equation \newvalley\ can be obtained by varying the following action,
$$S_{\rm NV} = S +S_\lambda, \quad 
	S_\lambda= -{1 \over 2 \lambda} \int d^4x 
	\left( {\delta S \over \delta\phi(x)} \right)^2_. 
\eqn\snewvalley$$ 
We introduce an auxiliary field $F(x)$ by adding the following term
to the above action, so that the four-derivative term is cancelled out.
$$S_{\rm F} = {1 \over 2 \lambda} \int d^4x 
\left(F(x) - {\delta S \over \delta\phi(x)} \right)^2_. \eqn\actionf$$ 
By varying the total action $S_{\rm NV}+ S_{\rm F}$, we obtain 
the following equations;
$$\eqalign{
	{\delta S \over \delta \phi(x) } - F(x) &= 0, \crr
	\int d^4x^\prime D(x, x^\prime)
		F(x^\prime) -\lambda F(x) &= 0. \crr}
\eqn\rewritten$$
The above set of equations is evidently equivalent to
the new valley equation \newvalley.
We also note that the solutions of the equations of motion
also satisfy the new valley equation with $F(x) = 0$.

\REF\bl{D.~E.~Brahm and C.~L.~Y.~Lee \journal\PR &D49 (94) 4094.}

The Euclidean action $S$ we study is the following;
$$S = \int d^4 x \ 
	\left[ {1 \over 2} 
	\left( \partial_\mu \phi\right)^2 + V(\phi)
	\right]_,
\quad
V(\phi) = {1 \over 2} \phi^2 (1- \phi)^2
	-\epsilon (4 \phi^3 - 3 \phi^4).
\eqn\action$$
The above potential has the false minima at $\phi =0$ and
the true minima at $\phi = 1$, regardless of the value of the
parameter $\epsilon ( \ge 0)$.
The energy density of the true vacuum is $-\epsilon$,
even for large $\epsilon$.
In this sense, this action defines a convenient model for study
of thick-wall bubbles as well as thin-wall ones.
It should also 
be noted that any quartic potential can be cast into the above form
by suitable rescaling of $x$ and redefinition of $\phi$.
In this sense, only the parameter $\epsilon$ is meaningful.
[Especially, the above form is related to that of
Ref.~\bl\ by a simple reparametrization.]

\def\dprime{^{\prime\prime}}
\def\sp{^\prime}
As the bounce solution is spherically symmetric, it is most
probable that the other configurations in the valley are also
spherically symmetric.
Thus, we shall confine ourselves to the study of spherically symmetric
configurations, 
$\phi(x) = \phi(\rho)$ and $F(x) = F(\rho)$, where
$\rho \equiv \sqrt{x_\mu^2}$.
The new valley equations \rewritten\ then lead to the following;
$$\eqalign{
	\phi\dprime + {3 \over \rho} \phi\sp
		- {d V\over d\phi} + F &= 0, \crr
	F\dprime + {3 \over \rho} F\sp
		- {d^2 V \over d\phi^2} F + \lambda F &= 0, \crr}
\eqn\actualeq$$
where we note the derivatives with respect to $\rho$ by primes.
Just as in Coleman's treatise of the bounce equation,
the above equations can be thought as the set of Minkowskian equations
of motion of a particle in two dimensional space $(\phi, F)$ 
at ``time" $\rho$.
The linear differential terms of $\phi$ and $F$ act as friction terms.
The other terms are space-dependent forces.
The big difference is that now this force is not conservative.
Therefore, no simple energy argument is possible.

The New Valley equations \actualeq\ require four boundary conditions. 
They are as follows:
For the solution to be regular at the origin $\rho=0$,
we require boundary conditions $\phi\sp(0)=F\sp(0)=0$.
Also, the outside of the bubble has to be the false vacuum, so
$\phi(\infty)=F(\infty)=0$ have to be satisfied.

In solving \actualeq\ numerically, we have chosen to start at the origin
with boundary conditions $\phi\sp(0)=F\sp(0)=0$ and
adjust $\phi(0)$ and $F(0)$ so that $\phi(\infty)=F(\infty)=0$
are (approximately) satisfied.
We have done this calculation for $\epsilon = 0.25$.
\FIG\pot{Potential $V(\phi)$ for $\epsilon=0.25$.}
The potential for this value of $\epsilon$ is given in Fig.\pot.
Since the depth of the true vacuum is much more than that of the height
of the potential barrier, we expect that the bounce solution
is a bubble with thick wall.
\FIG\solutions{Shapes of the solutions of the new valley equation \actualeq.
The eigenvalues $\lambda_{1\sim5}$ of each lines are 
$\lambda_1 = 0.3$, $\lambda_2 = -0.2$, $\lambda_3 = -0.34$, 
$\lambda_4 = -0.25$, $\lambda_5 = -0.2$.
The broken line shows the bounce solution.}
In Fig.\solutions,
we show the numerical solutions.
We observe the following in this figure:
1) The new valley contains the bounce solution (broken line).
	This solution has thick-wall as expected.
	It has a negative eigenvalue $\lambda_3$, 
	which causes the instability.
2) As $\lambda \rightarrow 0-$, large bubbles are created.
	The interior of these bubble is the true vacuum, $\phi=1$. 
The latter property is especially notable: Even though the bounce solution
is a thick-wall bubble, we find that the valley contains large, thin-wall,
clean bubbles in the outskirts.

These thin-wall bubbles can be analyzed by extending the original
Coleman's argument:
The solution of the new valley equation \actualeq\ extremizes the
action $S_{\rm NV}+S_{\rm F}$. 
Using the first equation of \rewritten, we rewrite it as follows
for negative $\lambda$;
$$S_{\rm NV}+S_{\rm F}=S + S_\lambda, \quad
	S_\lambda = {1\over 2 |\lambda|}\int d^4x F^2. \eqn\react$$
Now consider a fictitious particle in a one-dimensional space
$\phi$ at time $\rho$.
In the first equation \actualeq, the auxiliary field $-F(\rho)$ 
acts as an ``external force" for this particle.
If the initial value $\phi(0)$ is sufficiently close 
to $1$ and $F$ is sufficiently small, $\phi$ remains close
to $1$ for a long time, until the friction term dies away.
When it finally rolls down the hill, it does so under the external force 
$-F(\rho)$.
If $-F(\rho)$ is just right, the particle approaches to
the top of the lesser hill $\phi=0$ asymptotically. 
Note that there is a major difference with Coleman's
argument here: When the roll-down occurs, it does so under
the external force, which is the sole source of the 
energy reduction, while in the bounce solution the timing of the
roll-down has to be such that the friction term is just right 
to take care of the extra energy.
[This does not prove the existence of the thin-wall solution, but 
Fig.\solutions\ shows that this in fact happens.]
Since $-F(\rho)$ is the stopping force, it deviates from zero only at the wall.
Therefore the second term in \react\ contributes positively {\sl only} 
at the wall.
If we denote the radius of the bubble by $R$, 
the actions $S$ and $S_{\lambda}$ are approximately written as the following;
$$\eqalign{S &= -\epsilon {\pi^2 \over 2} R^4 + S_1 2\pi^2 R^3, \crr
	S_\lambda &= {W_F\over 2 |\lambda|} 2\pi^2 R^3,} \eqn\actrad$$
where $S_1$ and $W_F$ are the numbers of O(1).
Taking the derivative of $S + S_\lambda$ in the above 
with respect to $R$, we find the radius of the
solution of the new valley equation to be
$$R_{\rm NV} = {3\over \epsilon}
	\left(S_1 + {W_F \over 2 |\lambda|} \right). \eqn\rad$$
Therefore, even when $\epsilon$ is not small enough to guarantee
the large radius, $|\lambda|$ can be small enough to do so.
This is what is causing the thin-wall bubble to be a solution
of the new valley equation.

The shape of the large bubble and its thin wall can be
examined in detail by looking at the 0+1 dimensional model,
since the friction term can be neglected.
We have thus analyzed the 0+1 dimensional model also.
We have found that its valley also contains the thin-wall 
true vacuum bubbles, just as in 3+1 dimensional space-time.
We have obtained the numerical value $W_F \simeq 0.2104$ from this analysis.
The radius seen in Fig.2 for $\lambda_{4,5}$ is in agreement
with \rad\ for this value of $W_F$.

\FIG\value{The action of the solutions of the new valley equation
as a function of the norm $|\phi|$.
The points with eigenvalues $\lambda_{1\sim4}$ correspond to the lines in 
Fig.2.}
Numerical value of the action $S$ is plotted in Fig.\value,
where the horizontal coordinate is the ``norm" of the
solution, $|\phi| \equiv \sqrt{\int d^4x \ \phi(x)^2}$,
which for large $R$ should be $\sim \sqrt{\pi^2/2}R^2$.
We see in this figure that the bounce solution denoted by 
its smallest eigenvalue, $\lambda_3$, is in fact at the maximum point of
the valley.
We have also compared the asymptotic expression of action $S$
in \actrad\ with numerical values of $S$ for large $|\phi|$ 
and have confirmed that
the action is in fact dominated by the volume and the surface terms.

\FIG\energy{The energy of the bubbles at the mid-section,
where $\partial \phi / \partial \tau = 0$.}

On the other side of the valley are small bubbles.
In order to see their role, we have calculated the energy $E$
of the mid-section of bubbles and plotted the
result in Fig.\energy.
As is well known, the bubble solution ($\lambda_3$) has $E=0$, so that
the energy conservation allows it to contribute to 
the tunneling from the false vacuum.
We find that bubble smaller than the bounce has $E > 0$, while
larger ones have $E < 0$.
This shows that the smaller bubble contribute to the
tunneling from states with higher energy than the false vacuum.

In summary, we have studied the new valley method 
and have shown that it is a very powerful tool 
for the analysis of the structure of the valley.
We have numerically examined the valley that contains the bounce solution.
We have found here that even when the bounce solution is a thick-wall bubble,
the valley contains thin-wall large bubble, which interior is the true vacuum.
Smaller bubbles are also identified and found to contribute
to the decay of the higher energy states.
This is a rather interesting possibility, which should be explored further.
We have also shown in this letter that, unlike the streamline method, 
the new valley can be obtained from the action
$S_{\rm NV} +S_{\rm F}$. 
This enables us to calculate valley configurations 
at given $\lambda$ with good accuracy either by numerical integration 
or by variational methods.
This method should be useful for other applications of
the valley method, such as the tunneling between degenerate vacua.
Details of this analysis and further results will be published
in near future.\Ref\future{H.~Aoyama and S.~Wada, Kyoto University
preprint, KUCP-0073 (1994).}

\endpage
\refout
\endpage
\figout
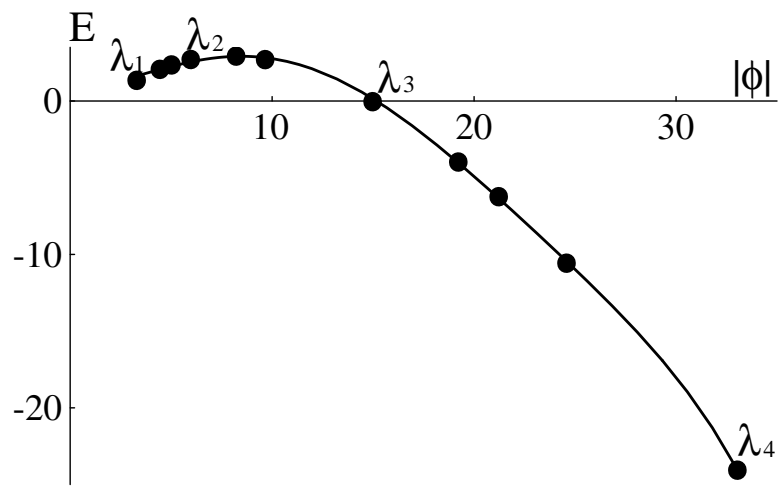
\end